\documentclass[aps,prb,twocolumn]{revtex4}
\usepackage{epsfig}

\newcommand{\kv}{\mbox{{\bf k}}}

\newcommand{\Av}{\mbox{{\bf A}}}
\newcommand{\Hv}{\mbox{{\bf H}}}
\newcommand{\Rv}{\mbox{{\bf R}}}

\newcommand{\lv}{\mbox{{\bf l}}}

\newcommand{\av}{\mbox{{\bf a}}}
\newcommand{\bv}{\mbox{{\bf b}}}
\newcommand{\gv}{\mbox{{\bf g}}}
\begin{document}
\title{Reentrant superconductivity in a strong applied field within the 
tight--binding model}
\author{Maciej M. Ma\'ska}
\email{maciek@phys.us.edu.pl}
\affiliation{Department of Theoretical Physics, Institute of Physics,\\ 
Silesian University, 40--007 Katowice, Poland}
\date{\today}

\begin{abstract}
It was suggested by Rasolt and Te\v{s}anovi\'c that the Landau 
level quantization in isotropic superconductors could enhance
superconductivity in a very strong magnetic field, above the upper 
critical field ($H_{c2}$). We derive a generalized Harper equation
for superconducting systems, and show that a
similar reentrant behavior appears in a lattice model, even though the 
Landau--level--structure is destroyed by the periodic potential in that 
case. Both the orbital and the Zeeman field--induced effects are taken 
into account. 
\end{abstract}

\maketitle

\section{Introduction}
There are two mechanisms responsible for a suppression of conventional 
superconductivity in an external magnetic field\cite{RT}: the Pauli pair 
breaking
and the diamagnetic pair breaking. First of them, the Pauli pair 
breaking, is connected with the Zeeman coupling. The magnetic field tends
to align the spins of the electrons forming the Cooper pair, and the singlet
superconductivity disappears at the Chandrasekhar--Clogston (CC) 
limit\cite{CC}.
However, this critical field for the majority of type--II systems is found 
to be above $H_{c2}$ determined by the orbital (diamagnetic) pair--breaking.
Especially, 
this effect is of minor significance in materials with low effective 
$g$--factor.
Another possibility is the superconductivity with nonhomogeneous
order parameter (the Larkin--Ovchinnikov--Fulde--Ferrell state\cite{LOFF}),
which can exist above the CC limit. One can also look for 
high--magnetic--field superconductivity in superconductors with triplet 
equal spin pairing.

The second effect, the diamagnetic pair--breaking, usually crucial in 
determining the upper critical field, is connected with the orbital 
frustration 
of the superconducting order parameter in a magnetic field. This frustration
enlarges the free energy of the superconducting state, and, when the magnetic
field is strong enough, the normal state becomes energetically favorable.
The orbital effect can be reduced in layered two--dimensional superconductors,
when the applied magnetic field is parallel to the conducting layers. Such a
situation has been analyzed theoretically\cite{Lebed} and recently observed 
experimentally in organic conductors\cite{Uji}.

However, it was shown that large values of the critical field are possible 
also in the
systems without two--dimensional layers, i.e., in systems where the
orbital effects are present. When describing a superconductor
within the Ginzburg--Landau-Abrikosov--Gor'kov theory \cite{GLAG}, one treats 
the magnetic field in the semiclassical phase--integral approximation, thus
neglecting the quantum effects of the magnetic field. This approximation 
is valid for relatively small fields, when $\hbar\omega_c \ll k_B T_c$ 
(or $\omega_c \ll
2\pi/\tau$ for large impurity concentration, where $\tau$ is the elastic
scattering time). In this regime, the number of occupied Landau levels is 
very large and                                                            
the energy spacing of them is very small, and therefore this discrete
structure is not observable. However, when the magnetic field increases,
the Landau-level degeneracy also increases, so the number 
of occupied levels decreases and one has to take it into account.
The inclusion of the Landau level quantization
in the BCS theory leads to reentrant behavior at a very high magnetic
field ($\hbar\omega_c \gg \epsilon_F$). Namely, when only the lowest Landau
level is occupied, $T_c$ is increasing function of $H$, limited only by 
impurity scattering and the Pauli pair breaking effect. 

The aim of this paper is to show that the reentrant behavior survives
in the presence of a strong periodic lattice potential. 

Weak, unidirectional periodic
potential removes (or, at least, modify) the Landau--level structure:
the levels are broadened (they form ``Landau bands'') and 
the degeneracy is lifted\cite{Langbein}. The width of a Landau band oscillates
as the magnetic field is tuned as a consequence of commensurability between the
cyclotron diameter and the period of the potential. It results in a 
magnetoresistance oscillations (Weiss oscillations\cite{Weiss}).
If the periodic potential is modulated in two 
dimensions, ``minigaps'' open in the ``Landau bands'', and the energy
spectrum of the system plotted versus the applied field composes
the famous Hofstadter butterfly\cite{Langbein,Hofstadter}, recently observed 
experimentally in the quantized--Hall--conductance measurement\cite{Albrecht}.
The same spectrum can be obtained in a complementary limit, when the lattice 
potential is strong (tight binding approach) and the field is weak. 
It is interesting that when the periodic potential does 
not lead to a scattering between states from different Landau levels, the 
eigenvalue equations in both the limiting 
cases are formally the same\cite{Langbein}. Of course the parameters
have a different physical meaning.

The simplest model for the case where a aplied field and a lattice
potential are present simultaneously, is commonly referred to as the Hofstadter
or Azbel--Hofstadter model \cite{Hofstadter,Azbel}. The corresponding 
Hamiltonian
describes electrons on a two--dimensional square lattice with the
nearest--neighbor hopping, in a perpendicular uniform magnetic 
field. The Schr\"odinger equation takes the form of a one--dimensional
difference equation, known as the Harper equation (or the almost
Mathieu equation) \cite{Harper,Hofstadter,Rauh}.
It is also a model for one--dimensional electronic system in two
incommensurate periodic potentials. The Harper equation has also
links to many other areas of interest, e.g., the quantum Hall effect,
quasicrystals, localization--delocalization phenomena
\cite{bellissard,guarneri}, the noncommutative geometry\cite{bellissard1},
the renormalization group\cite{thouless1,wilkinson}, the theory of 
fractals, the number theory, and the functional analysis\cite{last1}.

The Hofstadter model is useful in approach to the fundamental problem 
of the external magnetic field influence on the superconductivity.
Most of the works devoted to superconductors in the mixed state are based 
on the Bogolubov--de Gennes equations\cite{BdG}, particularly useful for
spatially inhomogeneous systems, e.g., for an isolated vortex\cite{wang} 
or a vortex lattice\cite{franz1}. However, in the regime 
$H_{c1} \ll H \ll H_{c2}$ we can neglect 
contributions to the spectrum from the inside of the vortex core 
(for $H \ll H_{c2}$
the distance between the vortices is large) and regard the magnetic 
field as uniform in the sample (for $H_{c1} \ll H$). We derive, under 
these assumptions,
a lattice model for the superconductor in applied field
(in the normal state such a system is described by the Hofstadter model).
In this paper we present a generalized Harper equation that describes the 
influence of magnetic field on the two--dimensional tightly bound electrons 
in the superconducting state.

\section{The model}

In analogy to Hofstadter's approach, we couple the magnetic field to the
system via the Peierls substitution\cite{pierls}, i.e., multiply the hopping 
matrix 
elements by a phase factor which depends on the field 
and on the position within the lattice. Thus, the vector--potential--dependent
hopping integral for sites $i$ and $j$ is given by
\begin{equation}
t_{ij}\left(\Av \right)= t
\exp\left(\frac{ie}{\hbar c} \int^{\Rv_{i}}_{\Rv_{j}}
\Av\cdot d\lv\right),
\end{equation}
where $t$ is the usual hopping integral.
We also include the Zeeman term. In effect, the BCS Hamiltonian has the form
\begin{eqnarray}
\hat{H}&=&\sum_{\langle ij\rangle,\sigma} t_{ij}\left(\Av \right) 
c^{\dagger}_{i\sigma}c_{j\sigma} 
%
%
+\sum_{i,\sigma}\left(\epsilon_\sigma-\mu\right)
c^{\dagger}_{i\sigma}c_{i\sigma} \nonumber \\
&&-\sum_{\langle ij\rangle}\left(\Delta_{ij} c^{\dagger}_{i\uparrow}
c^{\dagger}_{j\downarrow}
+\Delta^*_{ij} c_{i\downarrow}c_{j\uparrow}  \right),
\end{eqnarray}
where the Zeeman splitting is given by 
$\epsilon_\sigma=-\frac{1}{2}\sigma g\mu_{B}H$, $\sigma=1$ for spin up
and $-1$ for spin down,
$g$ is the Land\'e factor, and $\mu$ is the chemical potential.
Here, we have introduced the spin--singlet pair amplitude 
$\Delta_{ij}=\frac{V}{2}\langle c_{i\uparrow}c_{j\downarrow} -
c_{i\downarrow}c_{j\uparrow} \rangle$. The strength of the nearest
neighbor attraction $V$ is assumed to be field independent. The validity
of this assumption depends on the nature of pairing potential and the 
strength of the magnetic field. \cite{foot2}
For example, in the $t-J$ model 
$J_{ij}({\bf A})={4 t_{ij}({\bf A}) t_{ji}({\bf A})}/U$
is strictly field independent, since the change of the phase generated 
when an electron hops from site $i$ to $j$ and back, cancels out. 
Such an assumption has been also partially justified on the basis of 
antiferromagnetic--spin--fluctuation--driven superconductivity.{\cite{mmmm2}

Our starting point is a two--dimensional square lattice with basis vectors
$\av = (a,0,0)$ and $\bv = (0,a,0)$, immersed in a perpendicular,
uniform magnetic field $\Hv=(0,0,H)$. We choose the Landau gauge, 
$\Av = (0, Hx, 0)$. Since the vector potential $\Av$ is linear in $x$, the
translation corresponding to the vector $\av$ shifts the phase of the wave function.
This shift can be compensated by a gauge transformation, introducing
{\em magnetic translations}.
If the magnetic flux per unit cell, $\Phi$, is a rational multiple of 
the flux quantum $\Phi_0=hc/e$, i.e., if
\begin{equation}
\frac{\Phi}{\Phi_0}=\frac{p}{q},
\end{equation}
with $p$ and $q$ coprime integers, we can define {\em magnetic lattice},
with $q\av$ and $\bv$ as the basis of the {\em magnetic unit cell}. Such 
an enlarged unit cell is penetrated by $p$ flux quanta. Magnetic translations
corresponding to the {\em magnetic lattice} vectors 
($\Rv=nq\av+m\bv$, with $n,\:m$ -- integers)
commute with each other and with the Hamiltonian. If the system is of rectangular
shape with $L_x$ sites in the $x$ direction and $L_y$ sites in the $y$ direction,
and $L_x$ is a multiple of $q$, we can find eigenfunctions which diagonalize
the Hamiltonian and the magnetic translation operators simultaneously. Due to 
the absence
of translational invariance with vectors $m\bv$, vectors $\kv=(k_x,\:k_y)$
from the firs Brillouin zone $(|k_x|\le \pi/a,\ |k_y|\le \pi/a)$ are not 
good quantum 
numbers. Instead, we have to use vectors from a {\em magnetic (reduced)  
Brillouin zone} (MBZ), defined by $|k_x|\le \pi/qa,\ |k_y|\le \pi/a$, to enumerate 
the eigenstates. The Hamiltonian (1) in the momentum space can be written as
\begin{eqnarray}
\hat{H} & = & -t\sum_{{\bf k},\sigma} \left[ 2\,\cos \left(k_xa\right) 
c^\dagger_{{\bf k},\sigma} c_{{\bf k},\sigma} \right. \nonumber \\
& & \left. + e^{-ik_ya} c^\dagger_{{\bf k}-{\bf g},\sigma} 
c_{{\bf k},\sigma}
+ e^{ik_ya} c^\dagger_{{\bf k}+{\bf g},\sigma} c_{{\bf k},
\sigma} \right] 
\nonumber \\
& & + \sum_{{\bf k}\sigma}\left(\epsilon_\sigma-\mu\right)
c^\dagger_{{\bf k}\sigma}c_{{\bf k}\sigma}   
 - {\sum_{\bf k} \left( \Delta_{\bf k} c^\dagger_{{\bf k}\uparrow} 
c^\dagger_{{\bf -k}\downarrow} + {\rm H.c.} \right)},
\end{eqnarray}
where
\begin{equation}
\Delta_{\bf k}=\sum_{\bf k'} V_{\bf k,k'} 
\langle  c_{{\bf -k'}\downarrow} c_{{\bf
k'}\uparrow} \rangle ,
\end{equation}
and $\gv=\left(2\pi p/q,0\right)$.
In order to rewrite the above Hamiltonian as a sum over the MBZ we introduce
a multicomponent Nambu spinors
\begin{eqnarray}
C^\dagger_{\bf k} & = & \left( c^\dagger_{{\bf k},\uparrow},\
c^\dagger_{{\bf k}-{\bf g},\uparrow},\ c^\dagger_{{\bf k}-2{\bf g},\uparrow},\
\ldots ,\ c^\dagger_{{\bf k}-(q-1){\bf g},\uparrow},\ \right. \nonumber \\
&& \left. c_{-{\bf k},\downarrow},\ 
c_{-{\bf k}+{\bf g},\downarrow},\ c_{-{\bf k}+2{\bf g},\downarrow},\
\ldots ,\ c_{-{\bf k}+(q-1){\bf g},\downarrow} \right).
\end{eqnarray}
Then Eq. (4) can be written as
\begin{equation}
\hat{H} = {\sum_{\bf k}}' C_{\bf k}^\dagger H_{\bf k} C_{\bf k},
\end{equation}
where the prime denotes summation over the MBZ and $H_{\bf k}$ has a block structure
\begin{equation}
H_{\bf k} = \left( \begin{array}{c|c}
\ \ \hat{\bf T}_{{\bf k}\uparrow} \: \ & \hat{\bf \Delta}_{\bf k} \\ 
\hline 
\hat{\bf \Delta}^*_{\bf k} & -\hat{\bf T}_{-{\bf k}\downarrow}
\end{array} \right).
\end{equation}
The diagonal blocks describe noninteracting lattice fermions under the influence of 
magnetic field, and have the form similar to that derived by Hasegawa 
{\em et al.}\cite{hasegawa}
\begin{equation}
{\bf\hat{T}}_{{\bf k},\sigma}=-t \left( \begin{array}{cccccc}
M_{0,\sigma} & e^{ik_y} & 0 & \cdots & 0 & e^{-ik_y} \\
e^{-ik_y} & M_{1,\sigma} & e^{ik_y} & 0 & \cdots & 0 \\
0 & e^{-ik_y} & M_{2,\sigma} & e^{ik_y} & \cdots & \vdots \\
\vdots & \vdots & \ddots & \ddots & \ddots & 0 \\
0 & \vdots & 0 & e^{-ik_y} & M_{{q-2},\sigma} & e^{ik_y} \\
e^{ik_y} & 0 & \vdots & 0 & e^{-ik_y} & M_{{q-1},\sigma} 
\end{array} \right),
\end{equation}
where $M_{n,\sigma}=2\,\cos\left(k_xa + n\gamma\right)+\epsilon_\sigma-\mu$,
$\gamma=|{\bf g}|=2\pi p/q$, 
and diagonal matrix $\hat{\bf \Delta}_{\bf k}$ represents the pairing 
amplitudes
\begin{equation}
\hat{\bf \Delta}_{\bf k}=\textrm{diag}\left(
\Delta_{\bf k},\: \Delta_{{\bf k}-{\bf g}},\:
\ldots ,\: \Delta_{{\bf k}-(q-1){\bf g}} \right).
\end{equation}
Diagonalization of Eq. (8) provides a set of eigenenergies 
$\left\{{\cal E}_{{\bf k},i}\right\}$, where $i$ enumerates $2q$ 
values corresponding to a given ${\bf k}$ from the MBZ. 

The pairing amplitude in the presence of external magnetic field is determined 
self--consistently from the BCS--like equation
\begin{equation}
\Delta_{\bf k}=\frac{1}{2N}{\sum_{\bf k'}}'
\sum_{i=1}^{2q}
\frac{V_{\bf k,k'}\Delta_{\bf k'}}
{2{\cal E}_{{\bf k'},i}}\tanh\frac{{\cal E}_{{\bf k'},i}}{2k_B T} ,
\end{equation}
where $N=L_xL_y$ and the prime summation denotes again summation over the MBZ. 
In the following we restrict ourselves to the singlet pairing in the $s$--wave
channel $(\Delta_{\bf k}=\Delta)$, even though Eq. (11) is completely general.
Generally, this equation can be used, e.g., to analyze the 
magnetic--field--induced 
change of gap parameter symmetry\cite{krishana} or the upper critical field in 
the systems with the spin--triplet pairing. The latter area of application is 
especially attractive, since in these systems the Pauli
pair breaking mechanism is absent, and the upper critical field is expected
to be very high.\cite{JS} 

\section{Results}
\subsection{Orbital effects}

The transition lines $T_c(H)$, obtained in the absence of the Zeeman splitting
($g=0$), are presented in Fig. 1. 

\begin{figure}
\epsfxsize=8.5cm
\epsffile{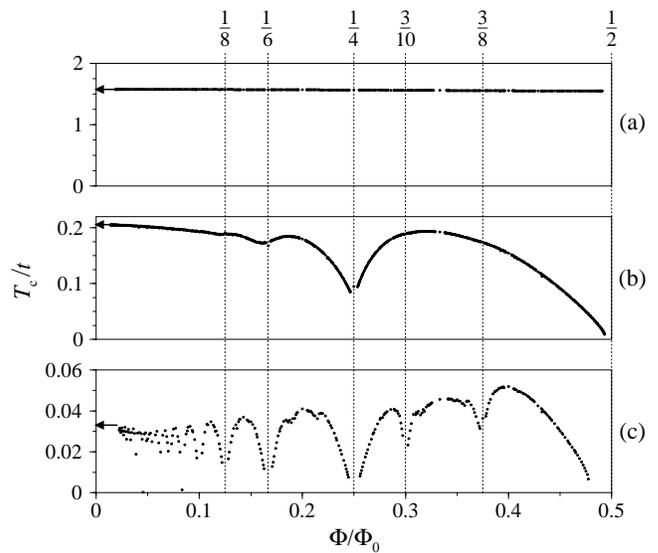}
\caption{Critical temperature $T_c$ as a function of $\Phi/\Phi_0$. Upper, 
middle, and lower panel shows results for $V/t = 7,\ 2$ and 1, respectively.
Small arrows indicate the transition temperature calculated from the usual 
BCS equations in the absence of magnetic field.}
\end{figure}

Note that for $a$ of the order of a few angstr\"oms, experimentally 
available magnetic fluxes are much less than $\Phi_0$. Consequently, 
these plots correspond to the region of extremely high magnetic field. 
The size of the 
Hamiltonian matrix $H_{\bf k}$, that has to be diagonized for all values of 
${\bf k}$ in each step of the iterative procedure, is $2q \times 2q$. 
Therefore, since the magnetic flux is proportional to $q^{-1}$, the 
proposed approach does 
not allow to carry out calculations for a small magnetic field. This is why the
transition lines in Fig. 1. start at finite magnetic field. 

For weak field thermal smearing and/or disorder induced broadening destroy the 
Hofstadter butterfly structure. In the absence of the lattice periodic potential
this regime corresponds to a classical limit, where the number of occupied Landau
levels is huge, and the Ginzburg--Landau description of the mixed state is
valid. In this regime, in accordance with the common feeling, superconductivity 
in the tight--binding system is suppressed by magnetic field, disappearing at 
$H_{c2}$\cite{mmmm2,mmmm1}. The corresponding transition line is presented 
in Fig. 2.

\begin{figure}
\epsfxsize=8.5cm
\epsffile{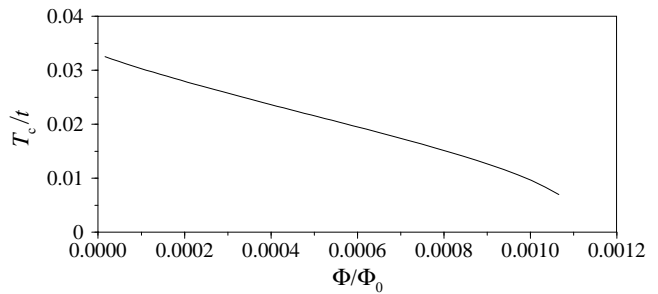}
\caption{Critical temperature $T_c$ as a function of $\Phi/\Phi_0$ in the low 
field regime for $V/t=1$. Results taken from Ref. \cite{mmmm2}} 
\end{figure}

The method used in Ref. \cite{mmmm2} does not work at low temperature and the 
present method does not work at weak field. Therefore, there is no crossover 
line from the low to high field regimes.

The transition lines for $1/2 <\Phi/\Phi_0 \leq 1$ can be obtained 
reflecting the lines presented in Fig. 1. around the line $\Phi/\Phi_0=1/2$,
 and $T_c(\Phi)$ is periodic on $\Phi_0$.
Both these properties reflect properties of the Hofstadter butterfly.
Of course, these unphysical results are valid only when the Pauli pair breaking
is neglected. The influence of the Zeeman splitting will be discussed later.
For strong pairing potential, comparable with the band width, the critical 
temperature in the reentrant regime is almost field independent (see Fig 1a.).
As $V$ is reduced, the influence of the nontrivial density of states becomes 
apparent. It was shown by Hofstadter\cite{Hofstadter}, that in normal state 
the Bloch 
band for $\Phi/\Phi_0=p/q$ is symmetric and broken up into $q$ distinct energy 
bands. In the half--filling case the Fermi level ($E_F$) is located in the 
center
of the (unperturbed) subband. Therefore, if $q$ is odd, $F_E$ points
to the singularity of the central subband (a remnant of the original van Hove 
singularity), whereas for even $q$ it is in the gap
between two subbands (In fact, for even $q$ these subbands touch itself at the
Fermi level). This is depicted in Fig. 3. 

\begin{figure}
\epsfxsize=8cm
\epsffile{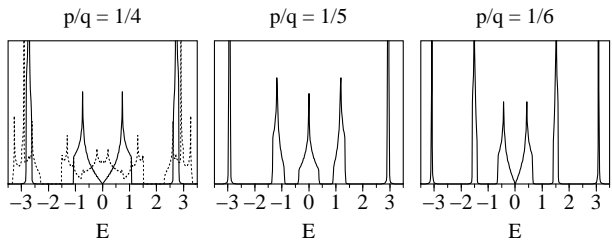}
\caption{Normal state density of states for different values of magnetic field.
The dotted line in the first panel represents density of states in a 
superconducting state for different ratio $p/q$ and for $\Delta=0.7t$.}
\end{figure}

The changes of the density of states
result in an oscillatory behavior of $T_c(H)$: $T_c$ approaches its maxima for
odd $q$ and is reduced for even $q$. Similar oscillations were predicted by
Rasolt and Te\v{s}anovi\'c \cite{RT} in a homogeneous system, where the 
Hofstadter spectrum is replaced by the Landau--level ladder. 

The superconductivity suppression is especially apparent for small and even 
$q$, when $V$ is comparable with the central--gap width. The smooth character 
of the
function $T_c(\Phi)$ close to $\Phi/\Phi_0=p/q$ and for small $q$ (e.g., 
close to the values $p/q=1/2,\ 1/3,\ 1/4$), seems counterintuitive, since
a tiny detuning of the magnetic field completely changes the spectrum.
For $p/q=1/2$ the spectrum consists of two subbands, whereas for $p/q=10/21$
there are 21 narrow subbands (see Fig.4). However, in spite of this difference,
the integrated densities of states, presented in Fig. 4c,
are almost the same.

\begin{figure}
\epsfxsize=8cm
\epsffile{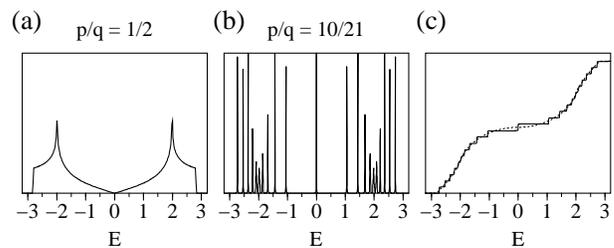}
\caption{Normal state density of states for $p/q=1/2$ (a) and $p/q=10/21$ (b).
(c) Comparison of integrated densities of states for $p/q=1/2$ (dotted line)
and $p/q=10/21$ (solid line).} 
\end{figure}

For larger $q$, the differences between $1/q$ and $1/(q+1)$ are smaller, and
consequently the distances between successive minima in the density of states 
decrease. For a strong pairing potential (and high $T_c$) 
 there is large number of subbands within a range of energy  
$\sim k_BT_c$ and then the amplitude of oscillations is strongly reduced.
On the other hand, for the weak potential (i.e., at low temperature), this 
irregular oscillations are visible even at low fields (cf. Fig. 1c).

\subsection{Zeeman splitting}

The previous discussion ignored the effect of Pauli pair breaking. We
consider it next. Since the Zeeman splitting is proportional to the 
magnitude of the
magnetic field and the orbital effect depends on the flux, we have to find
a relation between these two quantities. It can be done by using the
relation $t=\hbar^2/2m^* a^2$, where $m^*$ is the effective mass. 
Then the Zeeman splitting is given by $g\mu_B H=2\pi g^* \frac{p}{q}t$, 
where $g^*=g\frac{m^*}{m}$.\cite{Mo}

The inclusion of the Zeeman term results in a reduction of 
the phase space available for pairing. For strong pairing potential, when
the structure of the Hofstadter butterfly is hidden, it leads
to a monotonic reduction of $T_c$ with increasing magnetic field. Such a 
situation is presented in Fig. 5a.

\begin{figure}
\epsfxsize=8.5cm
\epsffile{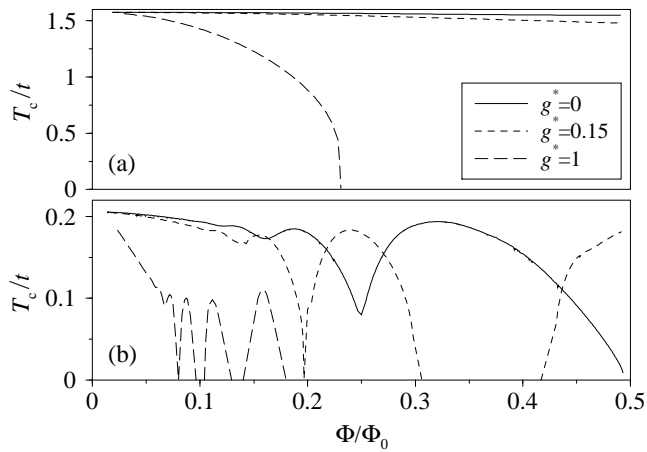}
\caption{Critical temperature $T_c$ as a function of $\Phi/\Phi_0$ in the 
presence of the Zeeman splitting for $V/t=7$ (a) and $V/t=2$ (b).}
\end{figure}

However, for smaller values of $V$, when $k_BT_c$
is comparable with the miniband (or minigap) widths, the situation
is more complicated. The Zeeman term leads to the splitting of each
of the minibands into a spin--up and spin--down minibands. To have nonzero
$T_c$ we need minibands of both types present close to the Fermi level.
As the magnitude of the splitting is proportional to the magnetic field,
$T_c$ will be an oscillatory function of the magnetic field. When the spin--up
and spin--down minibands overlap at the Fermi level, $T_c$ is strongly enhanced.
This mechanism may induce superconductivity in regions, where $T_c$ is zero or
close to zero in the absence of the Zeeman splitting (compare the solid and 
dashed lines in Fig. 5b). For example, for $\Phi/\Phi_0=1/4$ $E_F$
is located in the central minigap for $g^*=0$, whereas there is a singularity
at $E_F$ for $g^*=0.15$. The corresponding densities of states are
presented in Fig. 6.

\begin{figure}
\epsfxsize=8cm
\epsffile{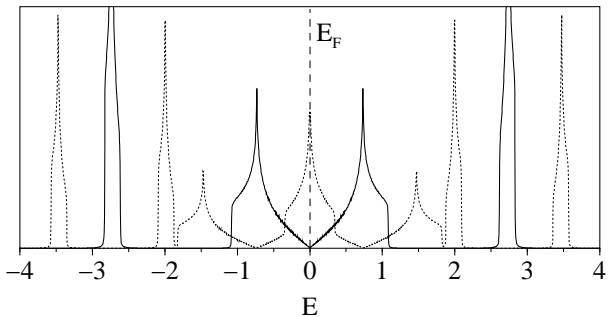}
\caption{Normal state densities of states for $\Phi/\Phi_0=1/4$. Solid and 
dashed lines correspond to $g^*=0$ and $g^*=0.15$, respectively.}
\end{figure}

\section{Discussion}

Let us comment on the possibility of observing the oscillatory behavior
of $T_c$ in real systems. Assuming lattice constant $a$=2\AA\ the magnetic 
field required to obtain $\Phi/\Phi_0\sim 1$ is $O(10^5)$T, which is obviously
too large. However, there are some possibilities to overcome this problem.
For example, it was recently shown\cite{3d}, that in some 
three--dimensional systems fractal spectra, like Hofstadter's butterfly, can 
be obtained for $\Phi/\Phi_0 \ll 1$. On the other hand, it is possible
to reach the needed increase of flux enlarging the lattice constant. 
Two--dimensional
superconducting wire network can be suitable for this task, since the 
magnetic field corresponding to $\Phi_0$ is about 1 mT for a network cell
of 1 $\mu{\rm m}^2$, and the system can be mapped onto a tight--binding one.
Another possibility is connected with the case, where the influence of
the modulation potential on the Landau--quantized 2D electron system may be
considered as a small perturbation. This situation is complementary to
the tight--binding case, but the energy spectrum is also obtained by solving
the Harper equation. Therefore, one can expect similar behavior of $T_c$
in 2D superconducting systems modulated in two dimensions. Again, since the
modulation period is larger than the lattice constant, the required values 
of $\Phi$ are well within the experimental accessibility.
 
\begin{acknowledgments}
The author is grateful to J\'ozef Spa{\l}ek for a fruitful disscussion. 
\end{acknowledgments}

\end{document}